\PassOptionsToPackage{expansion=false}{microtype}

\documentclass[]{vrk}

\usepackage[T1]{fontenc}
\usepackage{lmodern}
\usepackage{textcomp}

\usepackage{hyperref}
\usepackage{url}
\usepackage[most]{tcolorbox}
\usepackage{xcolor}
\usepackage{colortbl}
\usepackage{booktabs}
\usepackage{listings}
\usepackage{fancyvrb}
\usepackage{siunitx}
\usepackage{subcaption}
\usepackage{caption}
\usepackage{xcolor}
\usepackage{listings}
\usepackage[most]{tcolorbox}
\tcbuselibrary{listings,skins,breakable}

\newenvironment{boxedlst}{
	\begin{tcolorbox}[
		enhanced,
		breakable,
		colback=white,
		colframe=black,
		arc=2mm,
		boxrule=0.9pt,
		left=6pt,right=6pt,top=6pt,bottom=6pt,
		]
	}{
	\end{tcolorbox}
}

\title{
Design Conductor: An agent autonomously builds a 1.5 GHz Linux-capable RISC-V CPU
}

\author{The Verkor Team: Ravi Krishna, Suresh Krishna, David Chin}

\abstract{
Design Conductor (DC) is an autonomous agent which applies the capabilities of frontier models to build semiconductors end-to-end -- that is, from concept to verified, tape-out ready GDSII (layout CAD file). In 12 hours and fully autonomously, DC was able to \textbf{build several micro-architecture variations of a complete RISC-V CPU (which we dub ``VerCore'') that meet timing at 1.48 GHz\footnote{rv32i\_zmmul; using the ASAP 7nm PDK.}}, starting from a 219-word requirements document. The VerCore achieves a CoreMark score of 3261. For historical context, this is roughly equivalent to an Intel Celeron SU2300 from mid-2011 (which ran at 1.2 GHz). To our knowledge, this is the first time an autonomous agent has built a complete, working CPU from spec to GDSII. This report is organized as follows. We first review DC's design and its key components. We then describe the methodology that DC followed to build VerCore -- including RTL implementation, testbench implementation, frontend debugging, optimization to achieve timing closure, and interacting with backend tools. We review the key characteristics of the resulting VerCore\footnote{The VerCore RTL source and scripts necessary to rebuild the GDSII will be publicly available.}. Finally, we highlight how frontier models could improve to better enable this application, and our lessons learned as to how chips will be built in the future enabled by the capabilities of systems like DC.}

\date{\today}
\correspondence{\email{team@verkor.io}}

\begin{document}

\maketitle

\begin{figure*}[!htbp]
	\centering
	\begin{subfigure}[t]{0.48\textwidth}
		\centering
		\includegraphics[width=\linewidth, trim=0.5cm 2cm 0cm 2.5cm, clip]{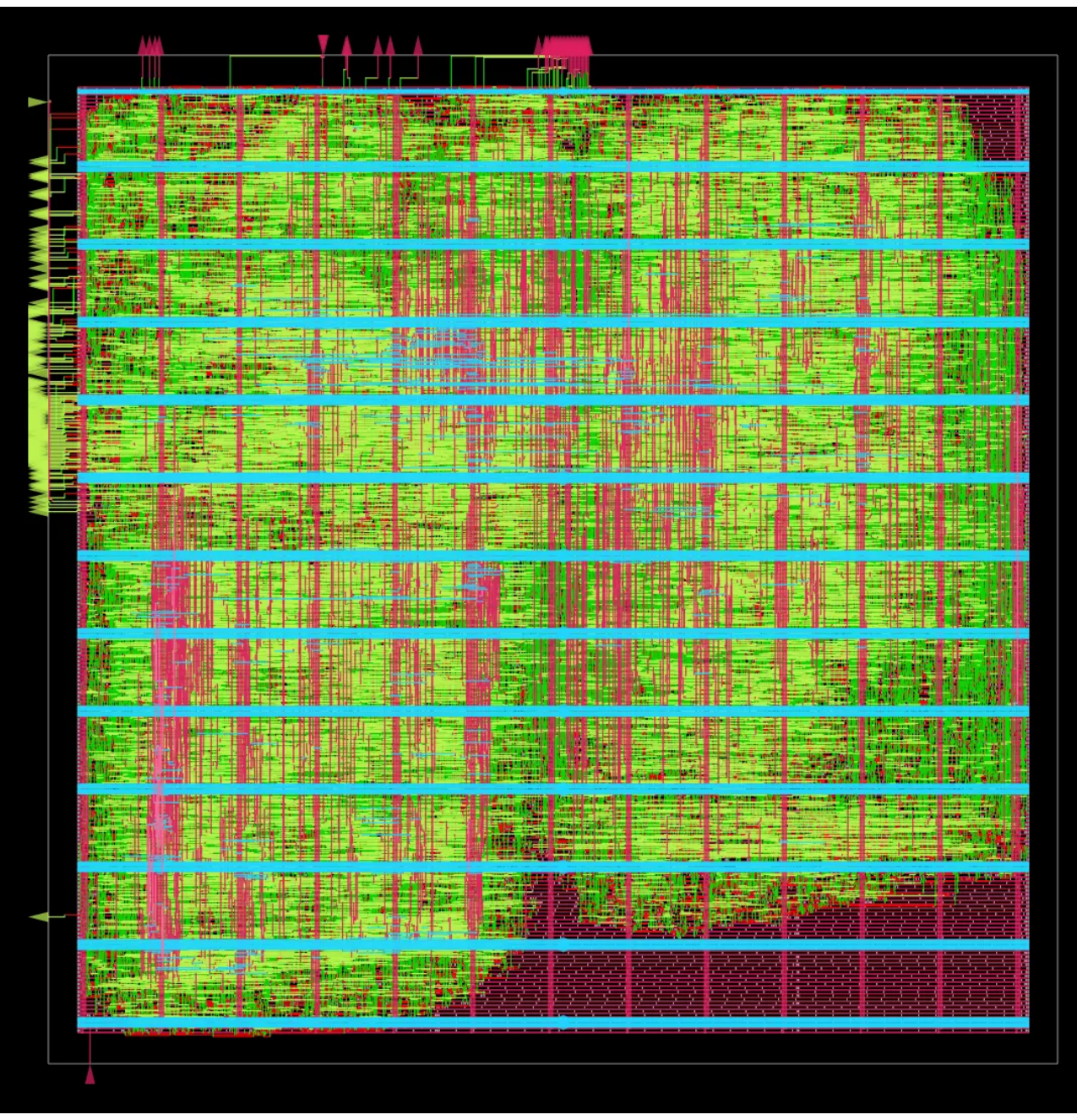}
		\caption{\textbf{VerCore design A (higher performance)}}
		\label{fig:vercoreA}
	\end{subfigure}\hfill
	\begin{subfigure}[t]{0.4765\textwidth}
		\centering
		\includegraphics[width=\linewidth, trim=0.68cm 2.32cm 0.5cm 2.5cm, clip]{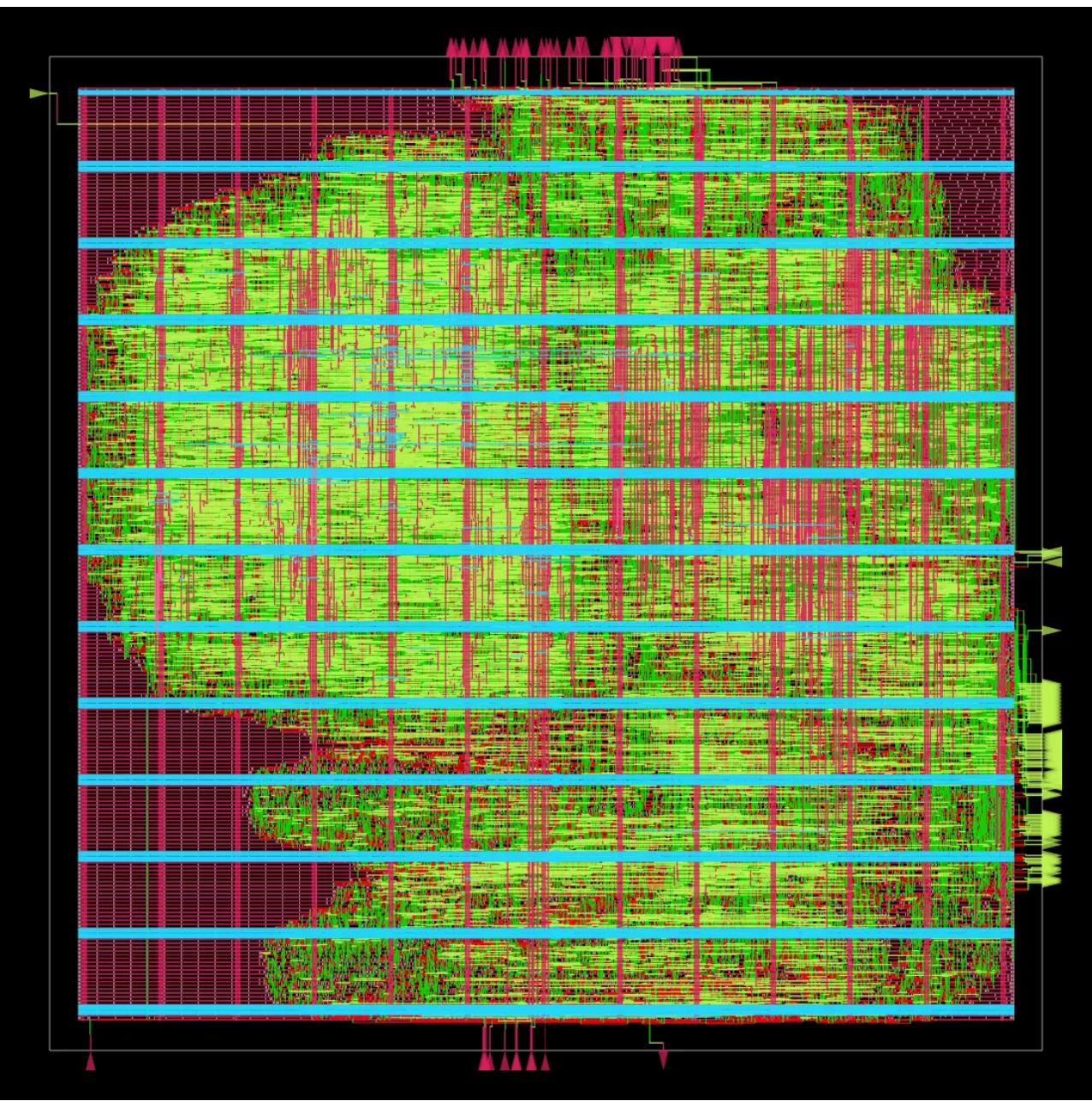}
		\caption{\textbf{VerCore design B}}
		\label{fig:vercoreB}
	\end{subfigure}
	\caption{\textbf{GDSII plots of VerCore designs (\SI{70}{\micro\meter} $\times$ \SI{70}{\micro\meter})}}
	\label{fig:vercore}
\end{figure*}

\section{Introduction}
\label{sec:intro}

Building a chip, whether from scratch or starting from an existing, older design,  is a highly time-consuming and expensive undertaking. It is widely understood that bringing a new leading-edge design to market costs well over \$400M and takes 18-36 months even with an engineering team numbering in the hundreds. A leading-edge chip design process encompasses many distinct steps, each of which can consume as much labor as a major software project. These include architecture definition, RTL implementation, testbench implementation and functional verification, frontend synthesis, place \& route, power estimation, and packaging. Further, commercially viable chips face several interlocking constraints that, in practice, require design iterations to meet. This consumes many additional months of time.

Chief among them is the requirement for extremely high functional test coverage -- that is, testing to ensure, with \textbf{\textit{very high}} confidence, that there are no ``bugs'' in the device's behavior. Because a single tape-out can cost tens of millions of dollars, ``fixing'' a bug in production is not an option. This contributes to the high cost of ``verification'', often estimated at over 50\% of total spend. Further, chips face multiple strict performance requirements, typically including at least on clock rate, power consumption, and silicon area (which affects production cost). The different types of simulation required for verification are also both slow in wall-clock runtime and expensive in terms of server-hours. Electronic Design Automation (EDA) tools are also highly configurable and require considerable expertise to operate in such a way as to achieve good end results for the design.

Due to these costs and challenges, many silicon markets are served by few vendors, with startup entry rarer than in software industries. Yet more potential designs are considered too low-volume to be worth a dedicated chip at all. And it typically takes many years and great engineering cost for the latest technologies to reach ultimate consumers.

Long-running autonomous AI agents present a promising opportunity to change this paradigm. To be fully effective in accelerating the design process, and avoid falling prey to Amdahl's law, such agents must address the entire problem -- all the way to tape-out ready GDSII \cite{gds1978gds}. The remainder of this report describes how Design Conductor (DC) works to do this.
\section{Design Conductor Architecture}
\label{sec:dc_arch}

This section reviews the key capabilities of DC and the architecture and infrastructure that supports them.

\subsection{Key Capabilities}

We list some of the key capabilities that DC was built to realize.

\subsubsection{Stable Long-horizon Execution}

As described in section \ref{sec:intro}, chip design is a complex task with many sub-components. DC has to be able to make progress towards a goal -- a functionally correct, high-performance design -- across many tens of billions of tokens consumed. This goal is not unitary but is rather a combination of several different design objectives  (power, performance, and area a.k.a PPA, functional constraints, and architecture input). DC must remember and meet all of these.

\subsubsection{Context Management}

DC must provide the underlying LLMs that it uses the information required for them to make good decisions. It must also carefully manage use of the limited context window, not only to avoid overflows but also to maximize quality.

\subsubsection{Technical Mastery}

LLMs are remarkable in their deep knowledge across many fields, and this is one way in which they arguably exceed human capabilities already. However, chip design requires extremely deep knowledge in certain specific areas. For example, master CPU designers understand certain ``tricks'' and ``recipes'' to achieve good performance. To be useful, DC must achieve a level of knowledge high enough that it can work smoothly with field experts. It must understand what it takes to achieve high performance across many types of designs.

\subsubsection{Correctness \& Verification}

``Vibe chip design'' is not an option when shipping millions of units. DC has to deliver a verifiably correct design.

\subsubsection{Balancing Exploration \& Speed}

The design space for chips is vast. DC has to be able to explore that space, while adhering to the user's instructions, in order to achieve the optimal performance. At the same time, DC must avoid going down a ``rabbit hole'' and failing to accomplish the overall objective in a timely manner. This requires managing search and exploration in a disciplined manner.

\subsubsection{End-to-end operation}

The most expensive and painful part of the human chip design process is a last-minute RTL change required to make timing (clock rate) targets or to fix a ``corner case'' functional bug right before tape-out. That is for obvious reasons: it requires ripping up at least some of the prior design effort, and introduces a risk of adding even more bugs. DC has to perform the same operations as it works to build a design, and it must do this while maintaining the necessary context and memory of its previous work.

\subsubsection{Infra}

Large-scale chip design is highly hardware resource intensive. VCD trace files used for debugging can be easily hundreds of gigabytes, and EDA tools use large amounts of DRAM to optimize designs during synthesis, placement, and routing. And DC may requires many subagent instances working together to accomplish its task in a timely manner. This means that DC's supporting infra has to be world-class in scalability and, importantly, reliability.

\subsection{Architecture}

\begin{figure*}[!htbp]
	\centering
	\includegraphics[width=1.2\textwidth, trim=4cm 8.5cm 4cm 2cm, clip]{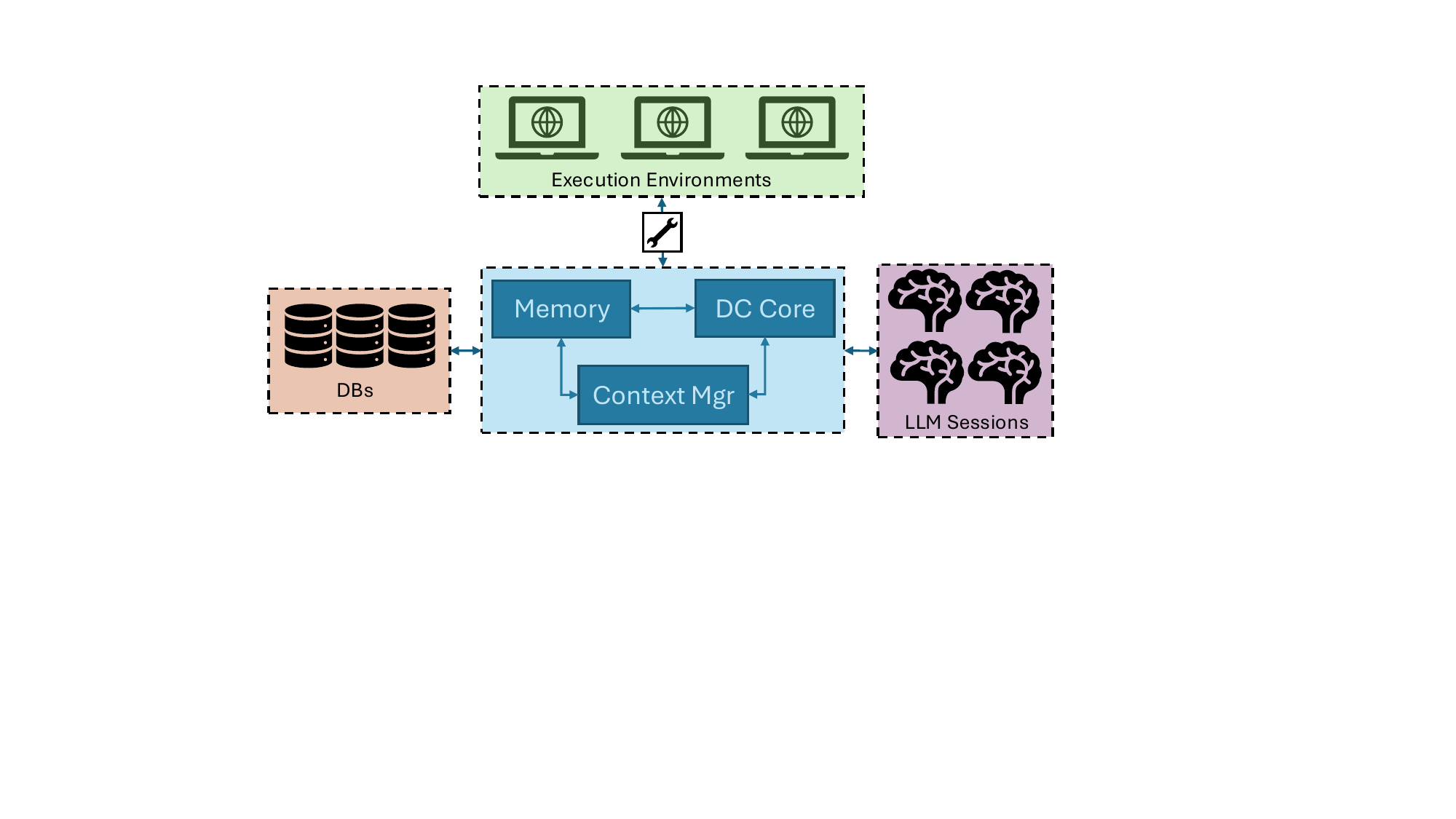}
	\caption{\textbf{High-level overview of Design Conductor's Architecture}}
	\label{fig:dc_arch}
\end{figure*}

We present a high-level architecture overview in figure \ref{fig:dc_arch}.

DC is a scalable, cloud-based application which runs on a distributed file system. LLM sessions are managed by worker servers, all of which sync to a central DB. These sessions connect to tool servers which exist in one or more execution environments, which can be VMs or containers. A context management module monitors and controls overall usage of the context window for the various sessions that are in-flight at any given time. In general, only the \texttt{Bash}, \texttt{Edit}, and \texttt{Subagent} tools are needed, although customized versions of these and additional tools can be used to improve quality. Subagents and higher-level algorithms (such as evolutionary) are managed by a top-level \texttt{DC Core} module, which interacts with lower-level LLM sessions.

Specific knowledge is provided to DC via a dedicated knowledge base. This knowledge base is contained within the main memory system. Memories exist indefinitely and are managed fully autonomously. DC makes use of this memory when it onboards itself onto a new codebase, or when it ingests requirements provided to it by users. This memory is also critical to ensuring that DC meets all requirements on the design requested by users, and that the design it is building meets all correctness requirements. A single DC ``instance'' is dedicated to one customer's design, such that no code, memories, or any information crosses between customers.

The actual design of these modules is proprietary and not discussed further in this report.
\section{Method}
\label{sec:method}

\subsection{Inputs to DC}

The only actual user input to DC was the following document:

\begin{tcolorbox}[colback=white,
	colframe=black,
	coltext=vrkblue,
	breakable,
	fontupper=\scriptsize\fontfamily{pcr}\selectfont]
		\textbf{VerCore RISC-V Design}
		\\ \\
		\textbf{Requirements Overview}
		\\ \\
		Your task is to build VerCore, a RISC-V CPU core that supports RV32I and ZMMUL, with the following hardware interfaces, all synchronous to a master clock:
		
		* Instruction cache interface (32-bit datapath)\\
		* Data cache interface (32-bit datapath)\\
		* Other interface signals: clock input to core, reset\_n input to core, asserted low.\\
		
		VerCore should implement a simple 5-stage pipelined design, in-order, single-issue of course.
		\\ \\
		DO NOT support compressed instructions.
		\\ \\
		Implement the register file as flip flops. This allows register reads to happen any time during the cycle, but writes happen at the next rising clock edge.
		\\ \\
		You need to achieve a CPI <= 1.5. Your overall goal is to maximize your design's score on CoreMark. Aim for a clock rate of 1.6 GHz.
		\\ \\
		You are responsible for both the RTL and the physical design. You should use the OpenROAD flow scripts to generate final GDSII output, along with area and timing information for this design. You should use the ASAP7 platform/PDK.
		\\ \\
		Assume that input signals will be valid 70\% into the clock cycle. Make sure output signals are valid 20\% into the clock cycle.
		\\ \\
		\textbf{Testing}
		
		You have access to Spike, the RISC-V ISA simulator. Use this to build a cycle-by-cycle integration test and verify that the behavior of your module matches that of Spike.
\end{tcolorbox}

DC was also given access to Spike \cite{spike}, the RISC-V ISA simulator, the RISC-V ISA \cite{waterman2011risc} and ASM manuals, as well as the RISC-V GNU toolchain.

\subsection{Steps followed by DC}

\begin{figure*}[!htbp]
	\centering
	\includegraphics[width=\textwidth, trim=1cm 0cm 5cm 1cm, clip]{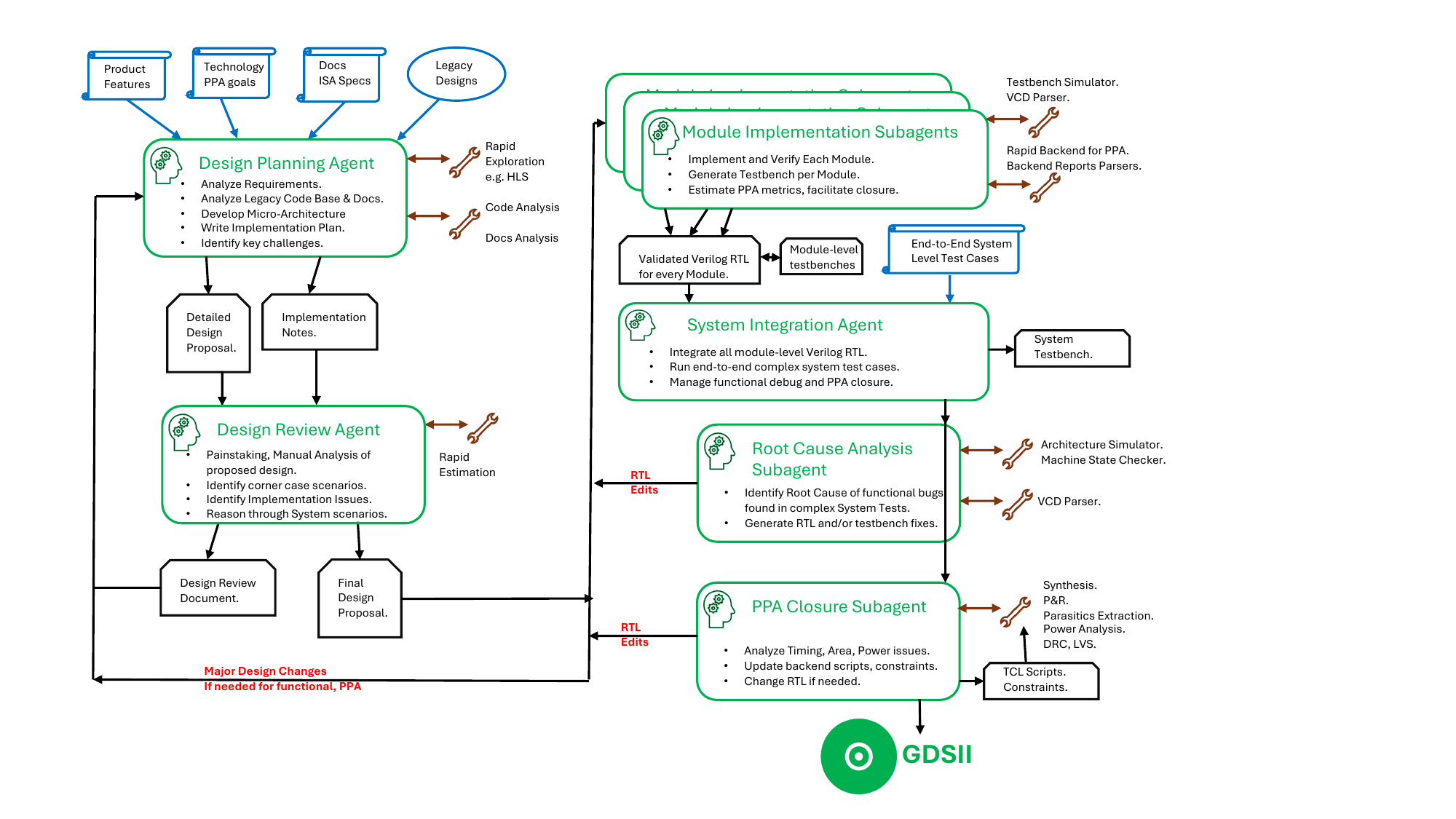}
	\caption{\textbf{Typical design process of Design Conductor. Note that for VerCore, there is no ``legacy design''.}}
	\label{fig:method_steps}
\end{figure*}

Figure \ref{fig:method_steps} illustrates the steps that DC went through to build the VerCore. \textbf{This process is ultimately under DC's control, and DC can customize or modify it and its execution to each design project. We simply provided the capabilities shown in the diagram as part of DC; their composition is determined by the \texttt{DC Core} module described in section \ref{sec:dc_arch}.}

DC starts with the input provided to it by the user. According to this input, its memory, and its knowledge, it generates an initial design proposal. An excerpt of this proposal is shown below. The design proposal is a ``living'' document, which DC updates as it fixes any functional or timing issues in its architecture. In fact we observed DC update its design based on feedback from final timing after place \& route.

\begin{tcolorbox}[colback=white,
	colframe=black,
	coltext=vrkblue,
	breakable,
	fontupper=\scriptsize\fontfamily{pcr}\selectfont,
	before upper={\setlength{\parindent}{0pt}\setlength{\parskip}{0pt}},] 
VerCore RV32I+ZMMUL Design\\
\\
Overview\\
- 5-stage in-order single-issue pipeline: IF, ID, EX, MEM, WB.\\
- PC reset to 0x2000. No compressed ISA. Register file uses flip-flops (sync write on posedge, async read).\\
- Cache interfaces per cache\_model.v: ic\_va/ic\_valid fetch (async read), dc\_va/dc\_valid/dc\_d\_out/dc\_byte\_en for loads/stores.\\
\\
Pipeline Stages (combinational stage modules, pipeline regs in vercore)\\
1) IF Stage (if\_stage)\\
- Holds PC in vercore. ic\_va = pc\_f. ic\_valid asserted when not in reset (held during stalls).\\
- pc\_plus4\_f = pc\_f + 4.\\
- PC update priority: reset -> branch/jump target -> stall hold -> pc+4.\\
- Branch/jump decision in ID drives pc\_next (1-cycle taken penalty via IF/ID flush).\\
\\
2) ID Stage (id\_stage)\\
- Decode opcode/funct3/funct7; generate control signals.\\
- Immediate generation for I/S/B/U/J types.\\
- Register file read (async). x0 always 0.\\
- Branch/jump resolved in ID using forwarded operands (from EX, MEM, WB). Branch compares for BEQ/BNE/BLT/BGE/BLTU/BGEU.\\
- JAL/JALR targets computed in ID (jal: pc+imm\_j; jalr: (rs1+imm\_i)\&{\textasciitilde}1). Jumps cause PC redirect and IF/ID flush. JAL/JALR writeback uses pc\_plus4.\\
- Branch instructions create no side effects (bubble into ID/EX).\\
\\
3) EX Stage (ex\_stage)\\
- ALU ops: ADD/SUB/SLL/SRL/SRA/SLT/SLTU/AND/OR/XOR.\\
- Address calc for load/store: rs1 + imm.\\
- AUIPC: pc\_ex + imm; LUI: imm.\\
- ZMMUL: mul/mulh/mulhsu/mulhu via riscv\_zmmul\_mul handshake. Mul result treated as ALU result; pipeline stalls while mul result pending.\\
\\
4) MEM Stage (mem\_stage)\\
- For loads/stores, drive dc\_* signals (gated by valid and global stall).\\
- Store byte enables/data:\\
* SB: dc\_byte\_en = 4'b0001 << addr[1:0]; dc\_d\_out = rs2 << (8*addr[1:0])\\
* SH: dc\_byte\_en = addr[1]?4'b1100:4'b0011; dc\_d\_out = rs2 << (8*addr[1:0])\\
* SW: dc\_byte\_en = 4'b1111; dc\_d\_out = rs2\\
- Loads: dc\_byte\_en=0. Select byte/half by addr[1:0] and sign/zero-extend per funct3 (LB/LBU/LH/LHU/LW). Load data computed combinationally in MEM for forwarding/WB.\\
\\
5) WB Stage (wb\_stage)\\
- Writeback selection: from MEM load data, ALU/mul result, or pc\_plus4 for JAL/JALR.\\
- regfile write occurs when wb\_reg\_write \&\& rd!=0; wb\_reg\_write itself is gated by rd!=0 to avoid x0 writes in spike\_checker. Signals exposed as u\_vercore.wb\_rd, u\_vercore.wb\_reg\_write, u\_vercore.wb\_data.\\
\\
Pipeline Registers\\
- IF/ID: pc\_id, instr\_id, pc\_plus4\_id, valid\_id.\\
- ID/EX: pc\_ex, pc\_plus4\_ex, rs1\_val/rs2\_val captured from forwarded ID-stage values (WB bypass), rs1, rs2, rd, imm, funct3/7, control (reg\_write, mem\_read, mem\_write, mem\_to\_reg, alu\_op, alu\_src, use\_pc/use\_imm, wb\_sel, mul\_en), valid\_ex.\\
- EX/MEM: alu\_result, store\_data, rd, funct3, control (reg\_write, mem\_read, mem\_write, mem\_to\_reg, wb\_sel), pc\_plus4\_ex, valid\_mem.\\
- MEM/WB: wb\_data, rd, reg\_write, valid\_wb.\\
\\
Forwarding\\
- EX stage forwarding (priority EX/MEM -> MEM/WB) for rs1/rs2.\\
- Store-data forwarding uses same sources for rs2 before latching into EX/MEM.\\
- ID stage branch/jalr forwarding from:\\
* EX stage ALU result (for immediately preceding ALU op)\\
* MEM stage forward value (load data if mem\_read else alu\_result)\\
* WB stage wb\_data\\
\\
Hazards / Stall / Flush\\
- Load-use hazard: if ID instruction uses rs1/rs2 and ID/EX is a load writing same rd, stall IF/ID + PC (hold), insert bubble into ID/EX.\\
- Branch/jalr hazard uses same load-use detection (stall when source depends on load in EX). Additional EX->ID forwarding resolves ALU dependencies without stall.\\
- Mul stall: when EX stage is mul and mul\_out\_valid=0, global stall holds all pipeline regs + PC. During global stall, dc\_valid and wb\_reg\_write are gated low to avoid repeat mem/reg events. When mul\_out\_valid=1, allow pipeline to advance and assert mul\_out\_ready to clear busy.\\
- Branch/jump taken in ID: update PC to target, flush IF/ID on update (insert bubble). Branch instruction itself is bubbled (no side effects). JAL/JALR continue for link writeback.\\
\\
	...remainder omitted...
\end{tcolorbox}

It will then review every aspect of this proposal. This review is ``manual'' and ``painstaking'', in DC's own words, to ensure the design is proper before implementation starts.  An excerpt of one such review for the multiplier unit design is shown below.

\begin{tcolorbox}[colback=white,
	colframe=black,
	coltext=vrkblue,
	breakable,
	fontupper=\scriptsize\fontfamily{pcr}\selectfont,
	 before upper={\setlength{\parindent}{0pt}\setlength{\parskip}{0pt}},]
==================================================\\
ZMMUL MULTIPLIER INTEGRATION DESIGN REVIEW\\
VerCore RISC-V Processor\\
==================================================\\
\\
DOCUMENT VERSION: 1.0\\
DATE: 2026\\
REVIEWER: Design Conductor (Subagent)\\
\\
==================================================\\
EXECUTIVE SUMMARY\\
==================================================\\
\\
REVIEW METHODOLOGY:\\
This review was conducted by manually and painstakingly walking through multiple\\
test scenarios, tracing cycle-by-cycle pipeline behavior, handshake protocol\\
operation, stalling mechanisms, and forwarding logic.\\
\\
SCENARIOS TESTED:\\
1. Simple multiply instruction\\
2. Back-to-back multiply instructions\\
3. Multiply followed by dependent ADD instruction (RAW hazard)\\
4. Multiply with load-use hazard\\
5. Multiply near a branch instruction\\
6. Edge case: Multiply with x0 destination\\
7. Edge case: Branch during multiply execution\\
\\
OVERALL ASSESSMENT: APPROVE WITH MINOR CONCERNS\\
\\
The design is fundamentally sound and should work correctly. However, there are\\
several areas that need attention:\\
\\
CRITICAL ISSUES FOUND: 0\\
MAJOR CONCERNS: 2\\
MINOR CONCERNS: 3\\
RECOMMENDATIONS: 5\\
\\
See detailed analysis below.\\
\\
==================================================\\
SCENARIO 1: SIMPLE MULTIPLY INSTRUCTION\\
==================================================\\
\\
TEST CASE:\\
Instruction sequence:\\
li   t0, 0x12345678    \# Load immediate (expands to lui + addi)\\
li   t1, 0x9ABCDEF0    \# Load immediate (expands to lui + addi)\\
mul  t2, t0, t1        \# MULTIPLY: t2 = (t0 * t1)[31:0]\\
li   t3, 0x242D2080    \# Expected result\\
\\
DETAILED CYCLE-BY-CYCLE TRACE:\\
\\
Assumptions:\\
- Pipeline starts empty\\
- No branch prediction, no cache misses\\
- Focus on the multiply instruction and surrounding context\\
\\
Let's trace starting from when the multiply enters the pipeline:\\
\\
Cycle 0: mul enters IF stage\\
IF: mul t2, t0, t1\\
ID: (previous instruction, e.g., li t1)\\
EX: (earlier instruction)\\
MEM: -\\
WB: -\\
\\
Pipeline state:\\
- mul instruction fetched\\
- No stalls yet\\
\\
Cycle 1: mul enters ID stage\\
IF: (next instruction, li t3)\\
ID: mul t2, t0, t1\\
EX: (previous instruction)\\
MEM: -\\
WB: -\\
\\
...remainder omitted...
\end{tcolorbox}

After this design phase, DC moves on to the actual module implementation. DC always builds per-module testbenches and fixes module functionality to ensure that these testbenches pass before proceeding.

At this point, DC focuses on integration testing. DC uses Spike to construct an overall \texttt{vercore\_tb.v} testbench which, given a RISC-V ELF, would run the test program on the DUT and confirm the design's architectural state and memory transactions match those reported by Spike. DC did this for many test programs, including MD5 and, ultimately, CoreMark itself.

\newpage
During this process, when any discrepancy with Spike is found, DC observes the conditions and examines the VCD file to debug the issue. It typically converts VCDs to CSV files and uses its inherent Python abilities to make processing easy:

\begin{boxedlst}
\begin{lstlisting}[caption={DC wrote this script to help it search through a VCD to debug an issue.\\}, language=Python]
import pandas as pd

df = pd.read_csv('vercore_tb.csv')

# Read expected register trace
with open('reg_trace.hex', 'r') as f:
    reg_trace = [line.strip() for line in f.readlines()]

print("=" * 80)
print("REGISTER WRITE COMPARISON")
print("=" * 80)

# Get actual writes
cols = [
    'time',
    'vercore_tb.u_vercore.wb_rd[4:0]',
    'vercore_tb.u_vercore.wb_data[31:0]',
    'vercore_tb.u_vercore.wb_reg_write',
    'vercore_tb.u_vercore.u_stage_if.pc[31:0]'
]

actual_writes = []
for idx, row in df.iterrows():
    if row['vercore_tb.u_vercore.wb_reg_write'] == 1:
        rd = int(row['vercore_tb.u_vercore.wb_rd[4:0]'], 16)
        data = int(row['vercore_tb.u_vercore.wb_data[31:0]'], 16)
        pc = int(row['vercore_tb.u_vercore.u_stage_if.pc[31:0]'], 16)
        time = row['time']
        if rd != 0:  # Skip x0
            actual_writes.append((time, rd, data, pc))

print(f"\nExpected writes (from reg_trace.hex):")
for i in range(min(10, len(reg_trace))):
    line = reg_trace[i]
    if len(line) >= 10:
        rd = int(line[0:2], 16)
        value = int(line[2:10], 16)
        print(f"  Write {i}: x{rd} = 0x{value:08x}")

print(f"\nActual writes (from VCD):")
for i, (time, rd, data, pc) in enumerate(actual_writes[:10]):
    print(f"  Write {i}: x{rd} = 0x{data:08x} (time={time}, PC=0x{pc:04x})")

print(f"\n" + "=" * 80)
print("MISMATCH ANALYSIS")
print("=" * 80)

# Compare
for i in range(min(len(actual_writes), len(reg_trace))):
    line = reg_trace[i]
    if len(line) >= 10:
        exp_rd = int(line[0:2], 16)
        exp_value = int(line[2:10], 16)
        
        if i < len(actual_writes):
            time, act_rd, act_value, pc = actual_writes[i]
            
            if exp_rd != act_rd or exp_value != act_value:
                print(f"\nMISMATCH at write {i}:")
                print(f"  Expected: x{exp_rd} = 0x{exp_value:08x}")
                print(f"  Got:      x{act_rd} = 0x{act_value:08x} (time={time}, PC=0x{pc:04x})")
                
                if exp_rd != act_rd:
                    print(f"  ERROR: Register address mismatch! Expected x{exp_rd}, got x{act_rd}")
                if exp_value != act_value:
                    print(f"  ERROR: Value mismatch! Expected 0x{exp_value:08x}, got 0x{act_value:08x}")
                break
        else:
            print(f"\nMissing write {i}: Expected x{exp_rd} = 0x{exp_value:08x}")
            break
\end{lstlisting}
\end{boxedlst}

\newpage
Using VCD analysis, DC then tracks down the root cause of the issue, proposes a fix, implements it, and tests again.

\begin{tcolorbox}[colback=white,
	colframe=black,
	coltext=vrkblue,
	breakable,
	fontupper=\scriptsize\fontfamily{pcr}\selectfont,
	before upper={\setlength{\parindent}{0pt}\setlength{\parskip}{0pt}},]
==================================================\\
TASK COMPLETION SUMMARY\\
==================================================\\
\\
TASK: Debug the register write mismatch error in the VerCore RISC-V CPU\\
\\
STATUS: COMPLETED\\
\\
==================================================\\
WHAT WAS DONE\\
==================================================\\
\\
1. Converted VCD to CSV for analysis\\
- Used vcd2csv.py to convert vercore\_tb.vcd to vercore\_tb.csv\\
- Extracted 26 rows and 403 signals\\
\\
2. Analyzed register trace expectations\\
- Parsed reg\_trace.hex to understand expected register writes\\
- Identified that write no. 3 should be x2 = 0x00003224\\
\\
3. Analyzed actual register writes from VCD\\
- Extracted all register writes from the simulation\\
- Found that write no. 3 was actually x5 = 0x0000300c\\
\\
4. Decoded instructions to understand the program flow\\
- Identified that PC=0x2008 has a JAL instruction jumping to 0x2020\\
- Instructions at 0x200c and 0x2010 should NOT execute\\
\\
5. Traced pipeline state during JAL execution\\
- Analyzed IF, ID, and EX stage contents over time\\
- Identified that branch\_taken is asserted correctly\\
- Found that pipeline flush is NOT working\\
\\
6. Identified root cause\\
- Pipeline flush logic is broken\\
- Instructions fetched after JAL are not being invalidated\\
- These instructions execute and cause wrong register writes\\
\\
7. Created detailed analysis document\\
- Includes timeline, evidence, and fix recommendations\\
- Visual representation of the bug\\
\\
==================================================\\
KEY FINDINGS\\
==================================================\\
\\
ROOT CAUSE:\\
Pipeline flush logic does not properly invalidate instructions when\\
branch/jump is taken. Specifically, when JAL at 0x2008 executes and\\
branch\_taken is asserted, the instruction at 0x200c (AUIPC x5) that\\
was speculatively fetched is NOT flushed and continues to execute.\\
\\
EVIDENCE:\\
- Time 85000: JAL in EX, branch\_taken=1, target=0x2020\\
- Time 95000: AUIPC x5 (from 0x200c) in EX - should have been flushed!\\
- Time 115000: x5 = 0x300c written - wrong! Should be x2=0x3224\\
\\
FIX REQUIRED:\\
\\
	...remainder omitted...
\end{tcolorbox}

It is this \textbf{verification-driven} approach that allows DC to arrive at a working design.

After all test programs pass the Spike-based testbench, DC moves on to PPA closure. DC reviews timing reports and uses this information to make RTL changes to the design. In this process, it figured out how to implement early forwarding in the ID stage, and it implemented a fast Booth-Wallace multiplier with 4 stages that are balanced and embody the most common forms of parallelism known to skilled designers.

DC can run indefinitely, but in this case we terminated its execution after a certain number of tokens consumed. The results as of that time are what we report in section \ref{sec:results}.
\section{Results}
\label{sec:results}

\subsection{Quantitative}

\begin{table}[h]
	\centering
	\begin{tabular}{l r}
		\hline
		\textbf{Metric} & \textbf{Value} \\
		\hline
		CoreMark \cite{coremark} score & 3261 \\
		Area (ex-cache) & \SI{2809}{\micro\meter\squared} \\
		Clock rate & \SI{1.48}{\giga\hertz} \\
		Technology & ASAP7 \\
		\hline
	\end{tabular}
	\caption{VerCore key metrics}
	\label{tab:metrics}
\end{table}

Table \ref{tab:metrics} shows the key quantitative metrics of the VerCore.

\subsection{VerCore Pipeline}

\begin{figure*}[!htbp]
	\centering
	\includegraphics[width=\textwidth, trim=0.5cm 1cm 1cm 1.5cm, clip]{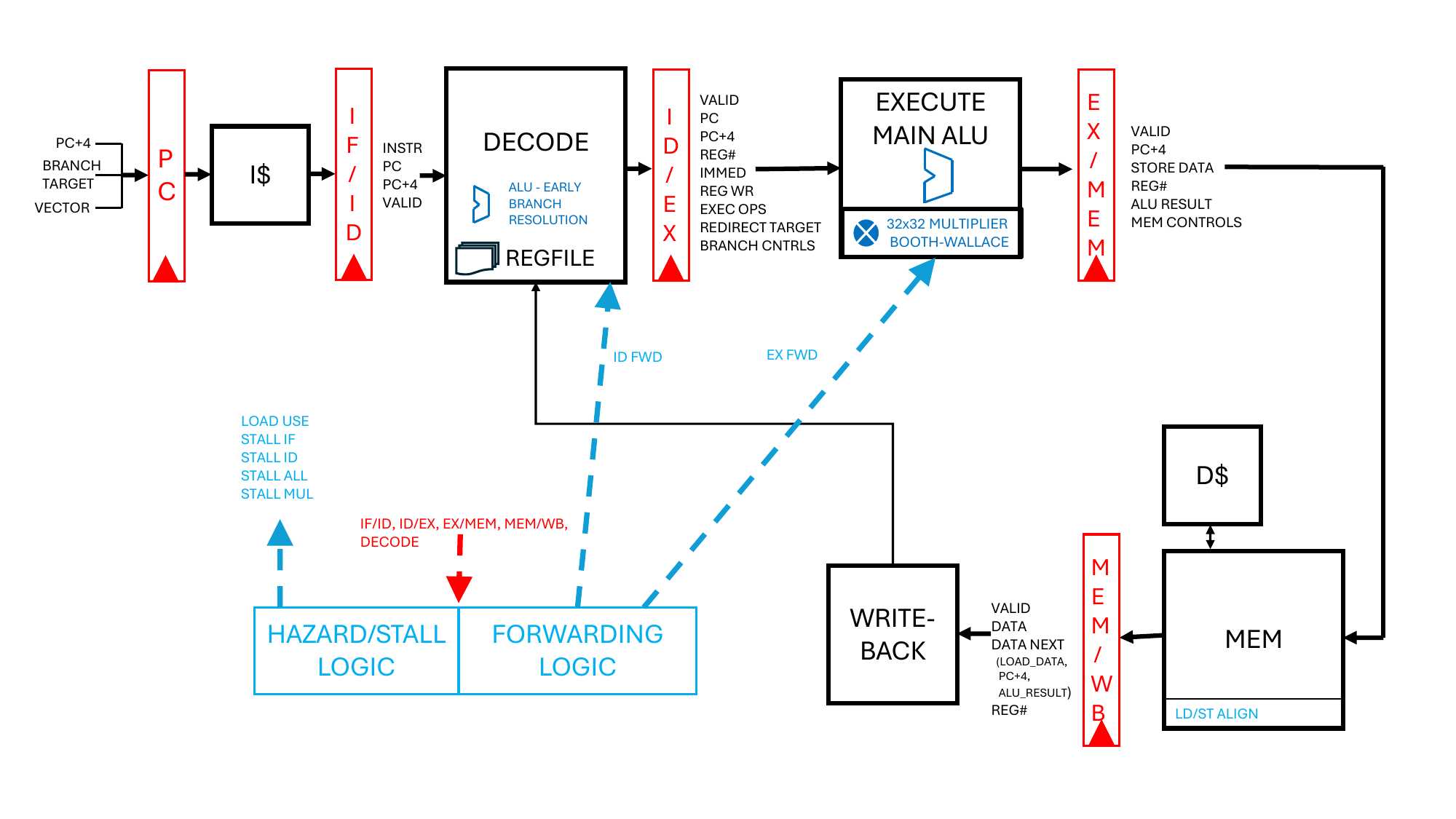}
	\caption{\textbf{VerCore pipeline diagram.}}
	\label{fig:pipe}
\end{figure*}

Figure \ref{fig:pipe} shows the pipeline of the final VerCore. In this work, DC generated multiple versions of the pipeline; the one we show in the figure achieves the highest performance of these. It features early branch resolution whenever possible, early forwarding, and a high-efficiency Booth-Wallace multiplier (which, on its own, clocks at 2.57 GHz). These attributes were discovered by DC and were not included in any input instructions (see \ref{sec:method}).

DC did not rely on ``guesswork''. Instead, DC did a full Verilog implementation of each variant (some with 2-cycle branch penalty, some with 1-cycle). DC fully implemented each variant down to GDSII. DC concluded that it could meet the clock rate target with a 1-cycle branch penalty design, even though that variant has a longer timing critical path that involves additional comparator logic. DC essentially re-discovered the critical path of the original MIPS 5-stage RISC CPU design, which also featured a 1-cycle branch penalty!
\section{Lessons for Frontier Models}
\label{sec:frontier_lessons}

We enumerate some of the ``LLM hang-ups'' we encountered in this work below. For these reasons, we believe it will remain critical to have experienced human architects guide systems such as DC.

\subsection{Architectural Reasoning}

One of the areas where base models require additional help is reasoning like an architect. We observed instances in which models made sub-optimal design choices that required a lot of tokens to ultimately optimize. For example, the forwarding implementations were often initially done in a manner that resulted in overly long critical paths. Only after the model observed timing results did it grasp the issue and fix this. This kind of knowledge often comes through experience for human designers.

DC also in some cases underestimates the complexity of work that is required to address certain issues. For example, in one case, when failing to meet timing, it initially tried making major modifications to deepen the pipeline, instead of looking for simpler explanations. This was a wasted exploration effort that consumed tokens unnecessarily, and could have been avoided with better architectural and engineering understanding.

\subsection{Understanding of RTL and Timing}

We observed instances in which models reason about Verilog, which is an event-driven language, as if it were sequential code. While we found that this did not impact DC's ability to achieve functional correctness, it made it more challenging for DC to debug timing issues. In one memorable instance, DC erroneously reasoned that reducing the number of dependent lines of code would result in a shorter critical path in the chip. These errors are ultimately corrected, because DC has access to actual timing reports from the tools, but they slow down DC's progress and consume additional tokens.

We posit that this is due to the heavy presence of software code in both LLM pre-training and post-training, and we expect that, as frontier labs recognize chip design as a more important application, this issue will be resolved.

\subsection{Specification Requirements}

We found that the input specification provided to DC has to be written in an extremely deliberate, tight, and verifiable/measurable manner. Without the CPI requirement in that document, for example, DC would sometimes generate a processor with significantly worse performance on branches and forwarding. With that line in the spec, DC would use a cycle counter in its testbench to compute its cycles per PC reported in the Spike trace to estimate CPI. In this way it was able to ensure it met the target.
\section{The Way Forward}

This section explains the authors' views as to how systems like DC can be scaled up to tackle commercial-complexity designs, and how human design teams can be structured to take full advantage of these new capabilities.

\subsection{Scaling}

We have found scaling to very large codebases -- e.g., with millions of lines of Verilog -- to not present any particular issue for DC. In testing on a codebase for a 13-stage OoO processor, DC was able to resolve functional and timing issues just as it was when dealing with the VerCore. This is due to the way that it structures codebase information in its memory.

The key challenge in addressing designs of this complexity is not in the mechanics of dealing with the codebase but rather in that DC needs to be operated by architects experienced in a given design area to achieve good results.

\subsection{The New Design Process}

\begin{figure*}
	\centering
	\includegraphics[width=0.95\textwidth, trim=3cm 3cm 0cm 0cm]{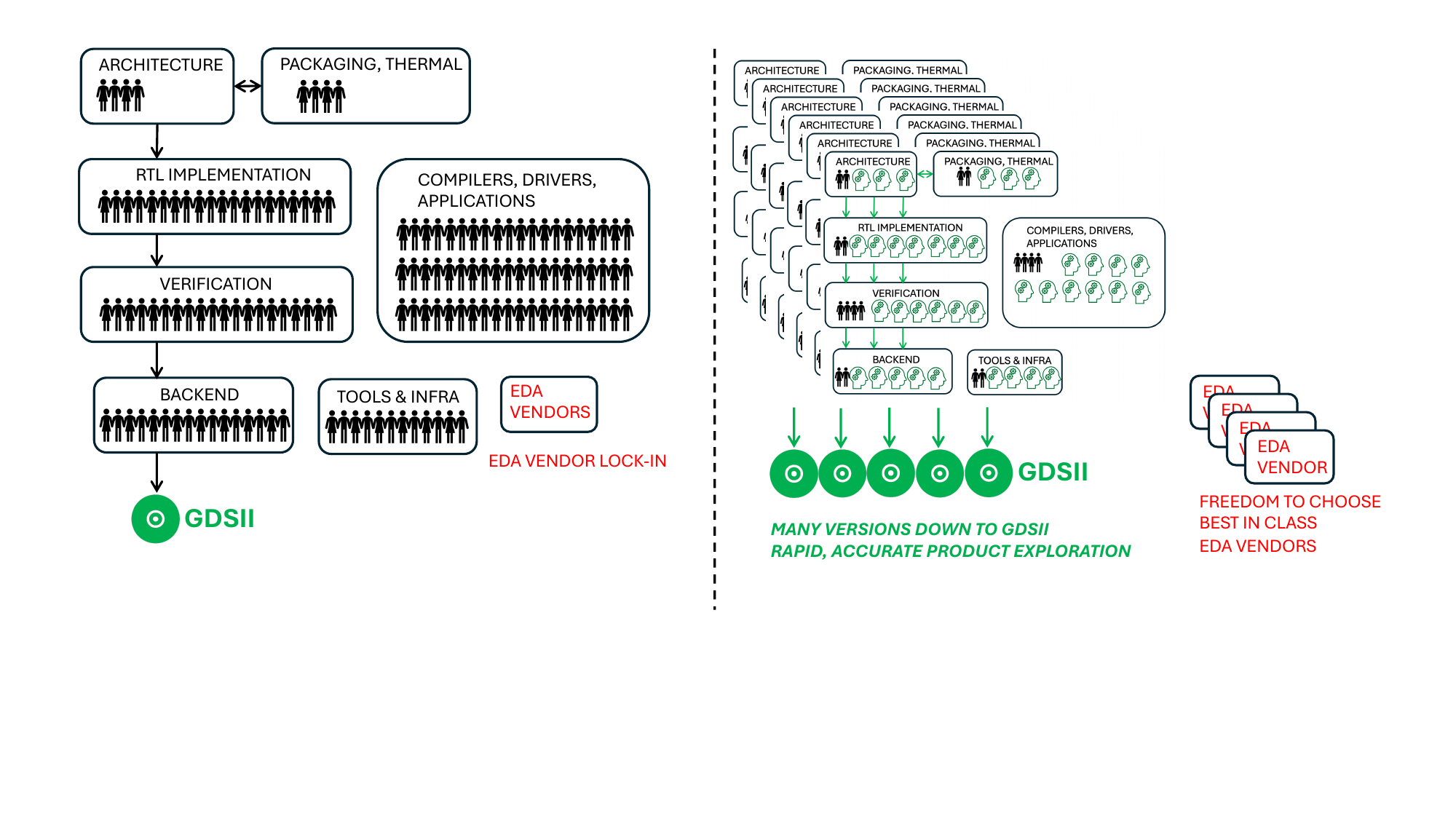}
	\caption{\textbf{An illustration of how teams might restructure to make the most effective use of DC. Many more designs and product ideas would be explored by a number of sub-teams, each capable of producing a single design from start to finish.}}
	\label{fig:team_impact}
\end{figure*}

Empowered with systems such as DC, teams that now comprise 100 or more people working on a single design will be able to simultaneously explore many different designs, architecture, and product ideas, each one from concept to GDSII. These teams will be able to tape out the most complex designs that exist in 3-6 months, instead of the current 18-36 months. The role of these experts will become to guide DC at an architecture and objective level to achieve the design result that they believe will succeed in the market -- with the ability to experiment without guesswork, and push for more aggressive cost and performance targets. Companies will also find many more addressable sockets in applications which were previously too low-volume to be served profitably.

One likely process change will be to front-load verification work, such that DC can be provided with some kind of integration test to guide its RTL implementation.

Senior engineers and master designers on such a team of the future will have many fewer ``tool-jockey'' responsibilities and will instead be relied on for their judgment and experience, with DC able to handle almost all other engineering work. DC's ability to manage tool interaction will also reduce tool switching costs and lock-in. Tool vendors will be able to focus on the quality of their algorithms, instead of expending effort on interface design and trying to ensure a simple environment for users.

\clearpage
\newpage

\bibliographystyle{plain} 
\bibliography{ref}

@misc{coremark,
	title        = {CoreMark EEMBC Benchmark},
	author       = {{EEMBC}},
	howpublished = {\url{https://www.eembc.org/coremark/}},
	note         = {Accessed: 2026-02-03}
}

@misc{spike,
	title        = {Spike, a RISC-V ISA Simulator},
	author       = {{RISC-V International}},
	howpublished = {\url{https://github.com/riscv-software-src/riscv-isa-sim}},
	note         = {Accessed: 2026-02-03}
}

@article{waterman2011risc,
	title={The risc-v instruction set manual, volume i: Base user-level isa},
	author={Waterman, Andrew and Lee, Yunsup and Patterson, David A and Asanovic, Krste},
	journal={EECS Department, UC Berkeley, Tech. Rep. UCB/EECS-2011-62},
	volume={116},
	number={2011},
	pages={1--32},
	year={2011}
}

@article{gds1978gds,
	title={GDS II Graphic Design System User’s Operating Manual},
	journal={CALMA, Nov},
	year={1978}
}

\end{document}